\documentclass[runningheads]{llncs}
\usepackage{subcaption}
\usepackage{graphicx}
\usepackage{wrapfig}
\begin{document}

\title{Metadata for Scientific Experiment Reporting: A Case Study in Metal-Organic Frameworks}
%
%
\author{Xintong Zhao\inst{1} \and
Kyle Langlois  \inst{2} \and
Jacob Furst  \inst{2}\and
Scott McClellan\inst{1} \and
Xiaohua Hu\inst{1}\and
Yuan An\inst{1} \and
Diego A. Gómez-Gualdrón  \inst{3}
Fernando J. Uribe-Romo \inst{2}
Jane Greenberg\inst{1}
}
\institute{Metadata Research Center, Drexel University, Philadelphia, PA, USA \\
\email{\{xz485,sm4522,xh29,ya45,jg3243\}@drexel.edu} 
\and 
Department of Chemistry, University of Central Florida, Orlando, FL, USA \\
\email{kylerlanglois@knights.ucf.edu,jfurst@knights.ucf.edu,fernando@ucf.edu}
\and
Chemical and Biological Engineering, Colorado School of Mines, Golden, CO, USA \\
\email{dgomezgualdron@mines.edu}
}



%
%
%
\maketitle              
\begin{abstract}
Research methods and procedures are core aspects of the research process. Metadata focused on these components is critical to supporting the FAIR principles, particularly reproducibility. The research reported on in this paper presents a methodological framework for metadata documentation supporting the reproducibility of research producing Metal Organic Frameworks (MOFs). The MOF case study involved natural language processing to extract key synthesis experiment information from a corpus of research literature.  Following, a classification activity was performed by domain experts to identify entity-relation pairs. Results include: 1) a research framework for metadata design, 2) a metadata schema that includes nine entities and two relationships for reporting MOF synthesis experiments, and 3) a growing database of MOF synthesis reports structured by our metadata scheme. The metadata schema is intended to support discovery and reproducibility of metal-organic framework research and the FAIR principles. The paper provides background information, identifies the research goals and objectives, research design, results, a discussion, and the conclusion.  

\keywords{FAIR,Research Reproducibility, Experiment Metadata, Metal-Organic Frameworks (MOFs), Synthesis Reporting}
\end{abstract}
\section{Introduction}

Metadata has been a key component of scientific research and data management even in the analog world. The significance of metadata has expanded with the growth in digital and networked technologies, data-driven science, and the evolution of data science and AI. This importance is central to the development and implementation of the FAIR principles, which are reviewed below. FAIR principles, together with open science, open data, and data sharing have also helped to highlight new metadata needs. One area in particular, is the need for metadata that specifically records research methods and procedures.

Metadata developments have emphasized resource discovery, access, and use, and aspects of provenance tracking and linking, interconnected with the linked data and the Semantic Web. Research methods follow general procedural frameworks, with some more specific than others. In the sciences research experiments, particularly those in the laboratory, are generally recorded in analog or digital lab notebooks. While an individual lab may have a recommended protocol or examples for reporting the experiment procedure, individual guidelines are over generalized or absent when researchers publish their results in scientific articles. Standardization in reporting of experimental research procedures in scientific publications is necessary to support reproducibility. The research presented in this paper considers this need with a specific materials science research case for the production of Metal Organic Framework (MOFs), which are produced through synthesis experiments.

The paper sections that follow provide background information on FAIR metadata for scientific research, metadata for research methods and procedures, and the MOF case. Next, the research goals, objectives, and research design and framework for identifying key MOF metadata entities is presented, followed by the results, a discussion, and, the conclusion.

\section{FAIR Metadata for Scientific Research}

Metadata is essential for supporting the  FAIR principles \cite{b1}, and making research data Findable, Accessible, Interoperable and Reusable. The goals embodied in the FAIR principles have been key to the open science \cite{b2,b3} and open data/data sharing movements \cite{b4}, even prior to the publication wide-spread global adoption of these principles. The FAIR principles along with good data management practices explain, in part, the wide availability of metadata standards for scientific research data. Key examples of globally used metadata standards for research data include the Data Document Initiative (DDI) \cite{b5,b6} for social science research data, the Darwin Core (DwC) \cite{b7,b8} for museum specimens, and the Ecological Metadata Language (EML) \cite{b9,ecoinformatics} for ecological research data. The Digital Curation Centre’s (DCC) Disciplinary Metadata directory
\cite{dcc}, and the Metadata Standards Catalog \cite{rdamsc} provide information on and links to these metadata standards (DDI, DwC, and EML) along with many other data-focused metadata standards. In addition to these disciplinary-wide directories, there are agency/domain focused initiatives, such as the United States Geological Survey (USGS) Metadata Creation guidelines \cite{Faundeen,usgs} and the United Nations Food and Agriculture (UNFAO) vocabulary services \cite{b12,b13}. These domain focused examples provide access to a suite of resources and tools to guide metadata generation and verification for research data and other digital resources. 

The spectrum of metadata standards for scientific research further includes an array of value standards (e.g., ontologies, taxonomies, vocabularies), and metadata applied to other aspects of a research undertaking, such as metadata reflecting access/usage rights \cite{b14}, metadata for physical samples \cite{b15}, or even research instrumentation and software \cite{b16,b17,b18}. And, finally, we cannot overlook the core component of scientific research metadata, specifically the well-known metadata that represents the entities associated with a publication citation, such as author name(s), title, publication date, persistent identifier (e.g., DOI) and metadata representing other publication properties.  This associated metadata is critical for research databases, citation indexes, and supporting FAIR. Overall, while the availability of metadata standards and metadata generation approaches  for documenting scientific research has rapidly advanced over the last two decades, it’s clear that less attention has been given to metadata specifically documenting the research methods and underlying procedures. This limitation motivates the work presented in this paper.

\section{Metadata for Research Methods and Procedures: A Case for MOFs}

Research is the drive for knowledge discovery. Research is shaped by the scientific method, with the goal that research should be verifiable \cite{b19}. Researchers pursue new knowledge by asking questions, testing hypotheses, measurement, experimentation, and other mechanisms. The field of research is shaped by research methods and procedures. Among commonly taught textbook methods are surveys, interviews, observational studies, anthropological approaches, and experimental design. Each of these methods involve a set of prescribed steps, and researchers frequently pursue mixed or multi-method approaches. For example, some researchers will provide details, indicating the steps, whereas others may tend to report the procedures at a more general level. 

Research has emphasized the need for more explicit metadata standards targeting research methods. For example, Chao \cite{b20} sought to identify metadata properties used to record methods across a set of eight schemes frequently used for reporting scientific research. Her results show explicit and implicit aspects, with four schemes having a mechanism for recording methods.  Among this set, only DDI and ThermoML (TML), indicated that the methods-related metadata properties were mandatory. Researchers focusing on paradata \cite{b21}, provenance and pipelines \cite{b22}, and reproducible computational research \cite{b23} underscore the importance of metadata in this area. Additionally, some research journals provide more concrete guidelines on reporting research procedures, although such guidelines are not an official standard, they provide a framework. To this end, some researchers also publish supplemental material that provides detail on the procedures.  

The developments highlighted here and Chao’s work, in particular, point to an interest in standardization of research reporting. This need has is an important goal for materials scientists working in the area of Metal Organic Frameworks (MOFs), the research area that presents the case for the work reported on in this research.  

Metal Organic Frameworks (MOFs) materials are a type of crystalline materials with open porous architectures with a predetermined structure \cite{yaghi2003}. MOFs  present an important class of materials as they have shown significant potential in various applications such as CO2 emission, catalysis and drug delivery. The large number of publications in the MOF area shows the diversity of research approaches. In addition, given the complexity of design space of MOF materials, a significant difference appears among synthesis experiment reports from researchers. Since no specific standards for experiment reports exist, the details from MOF synthesis experiments are not consistently reported. For example, some synthesis reports can be more detail-driven whereas some are more generalized; also, experimentalists could include different types of details in their reports. The current state of reporting severely limits researchers’ abilities to achieve FAIR, particularly reproducibility, and motivates research to  our goal to identify key metadata entities for standardizing MOF synthesis experiment reporting.

\section{Research Goals}

The goal of this study is to identify key  metadata components to facilitate experiment reporting, using Metal-Organic Framework (MOFs) as a case study. We present this work as a case study in that the results can contribute to larger arena of reporting scientific research. In this paper, we report our framework to design a metadata standard that structures synthesis experiments reported in research articles. As a result, this “codification” of synthesis recipes is beneficial to support discovery and reproducibility of metal-organic framework materials. 

To achieve this goal, we identified the following objectives:
\begin{itemize}
\item What are the key components in a synthesis experiment?
\item How to quantify the key components by pairs of entities and relations?
\end{itemize}

\section{Research Design and Framework}

We pursued the MOF case study method using natural language processing and interviews to gather domain expert feedback to develop the framework for metadata standard for MOFs synthesis experiment. Overall, our framework contains four main steps:

\begin{itemize}

 \item Identifying key components in MOF synthesis experiments and initialize metadata design
 \item Gather corpus of research articles and filter paragraphs related to synthesis procedure
 \item Apply the initial design of metadata to filtered synthesis paragraphs and send to domain experts for assessment
 \item Modification of metadata structure based on the feedbacks from domain experts
Detailed description of each step is shown below.
\end{itemize}

\subsection{Identifying key components in MOF synthesis experiments and initialize metadata design}

\begin{figure}
\centering
\begin{subfigure}[b]{0.8\textwidth}
   \includegraphics[width=1\linewidth]{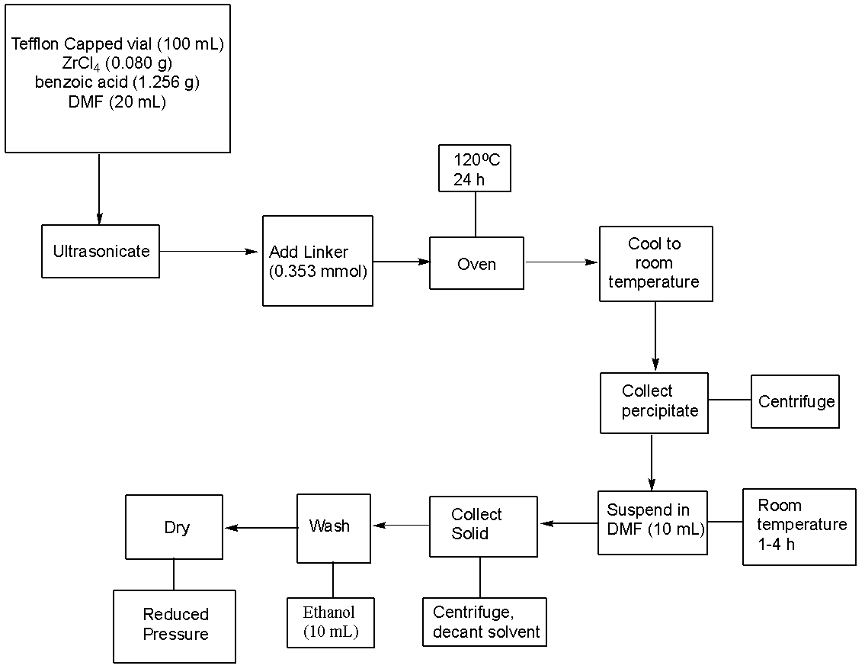}
   \caption{An example of MOF synthesis procedure}
   \label{fig1a} 
\end{subfigure}

\begin{subfigure}[b]{0.8\textwidth}
   \includegraphics[width=1\linewidth]{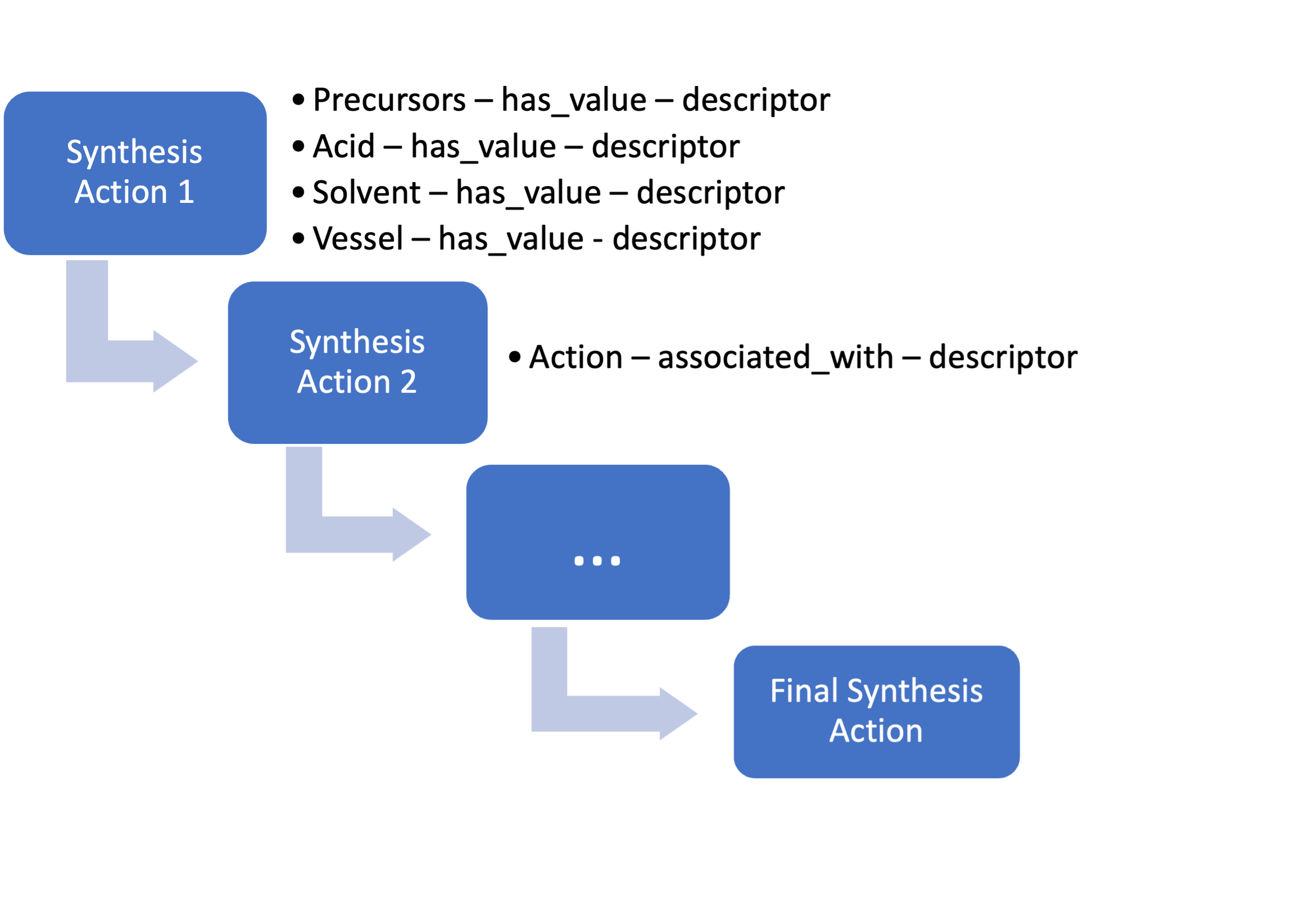}
   \caption{Metadata Design to Describe Synthesis Experiment for MOF materials}
   \label{fig1b}
\end{subfigure}

\caption[]{Identifying key components in MOF synthesis experiments and initialize metadata design}
\label{fig1}
\end{figure}

We conducted qualitative analysis to gain better understanding of domain-specific needs in order to design the metadata. Namely, we performed the following two steps: first, we reviewed previously established research about MOF article text mining \cite{b24,b25}; second, we arranged interviews with materials scientists to identify the critical components in synthesis experiments. Based on the results of two steps above, we quantified the synthesis procedure using the metadata structure demonstrated in figure \ref{fig1}. Figure \ref{fig1a} is an example of a real synthesis procedure to create a MOF material. Based on the findings from literature review and interviews with domain scientists, we quantify such synthesis procedure as a sequence of actions associated with specific attributes. The design of metadata is shown in figure \ref{fig1b}: the whole synthesis experiment is linked by synthesis actions (e.g., ultrasonicate, cool, dry), and specific details (e.g., acids, chemical composition/precursors, numerical values) are associated with actions. To this end, we design our metadata for synthesis reports as a set of entities linked by a set of relations. An initial list of key entity types is determined. We identified entity types that appeared in previous studies and incorporated them into our metadata design. The following entity types are included in the initial list: chemical composition(i.e., precursors), synthesis action and numeric descriptor.

In addition to the entity types, we also defined two relation types to link pairs of entities: (1) has\_value and (2) associated\_with. The entity relation pair plays an important role in the synthesis procedure of metal-organic frameworks. It can indicate links between two different entities; for example, when several synthesis actions (e.g., cooled, heated, ultrasonicated) and some descriptors (e.g., room temperature, 2 hours) are identified, we can link actions with their corresponding descriptor explicitly by assigning relations between them (e.g., cooled under room temperature). In this step, we initialized a list of entity and relation types to describe synthesis data.

\subsection{Gather corpus of research articles and filter paragraphs related to synthesis procedure}
After initializing our metadata to describe synthesis experiments, the next step will be to assess the initial design and make further modification. To this end, we decided to apply our initial metadata design to synthesis experiments reported in different research articles. To gather such a corpus consisting of MOF synthesis experiments, we collected 215,866 full research articles from the American Chemical Society (ACS) that are categorized under materials science study. Then we filtered MOF-related articles by matching their DOI numbers to DOIs listed in the Cambridge Structural Database (CSD), which is a major database registering crustal structures like MOFs. 

\begin{figure}
\centering
\includegraphics[width=0.9\textwidth]{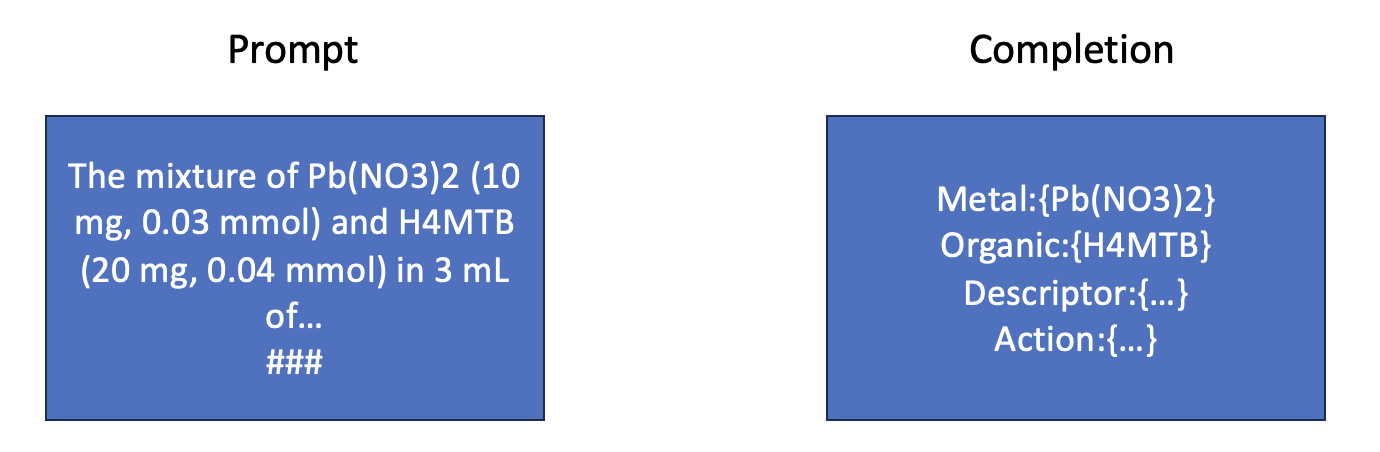}
\caption{An Example of Used Prompt for Named Entity Recognition} 
\label{fig4}
\end{figure}

After finding MOF articles, we used a few-shot learning method based on GPT (Generative Pre-trained Transformers) models to identify specific paragraphs that describe the synthesis procedure - we first manually identified a few synthesis paragraphs as examples, then applied the fine-tuned GPT models to all collected articles to identify synthesis paragraphs. We selected GPT model \textit{Curie} to perform the namend entity recognition. To fine-tune the \textit{Curie} GPT model, we applied prompt engineering based on the manually annotated examples. The fine-tuning process was handled by OpenAI API\cite{openai}. An example of prompt and completion structure for the few-shot learning is demonstrated in figure \ref{fig4}. Among all MOF-related articles, 2796 synthesis paragraphs were identified.

\subsection{Send sample paragraphs to domain experts for initial assessment}
Among synthesis paragraphs collected above, 15 paragraphs were randomly selected and sent to domain scientists to perform a first round assessment of the metadata schema. Two domain scientists were asked to annotate the 15 synthesis reports based on the metadata designed above. During the annotation process, they need to provide answers to the following questions:

\begin{itemize}

\item Is current metadata able to include all important information described in the experiment report?
\item If the current metadata does not capture all critical information from the synthesis steps, what should we add to the metadata?
\item Does current metadata design contain redundant components? If yes, then what should we remove from current metadata design?
\end{itemize}

\begin{figure}
\centering
\includegraphics[width=\textwidth]{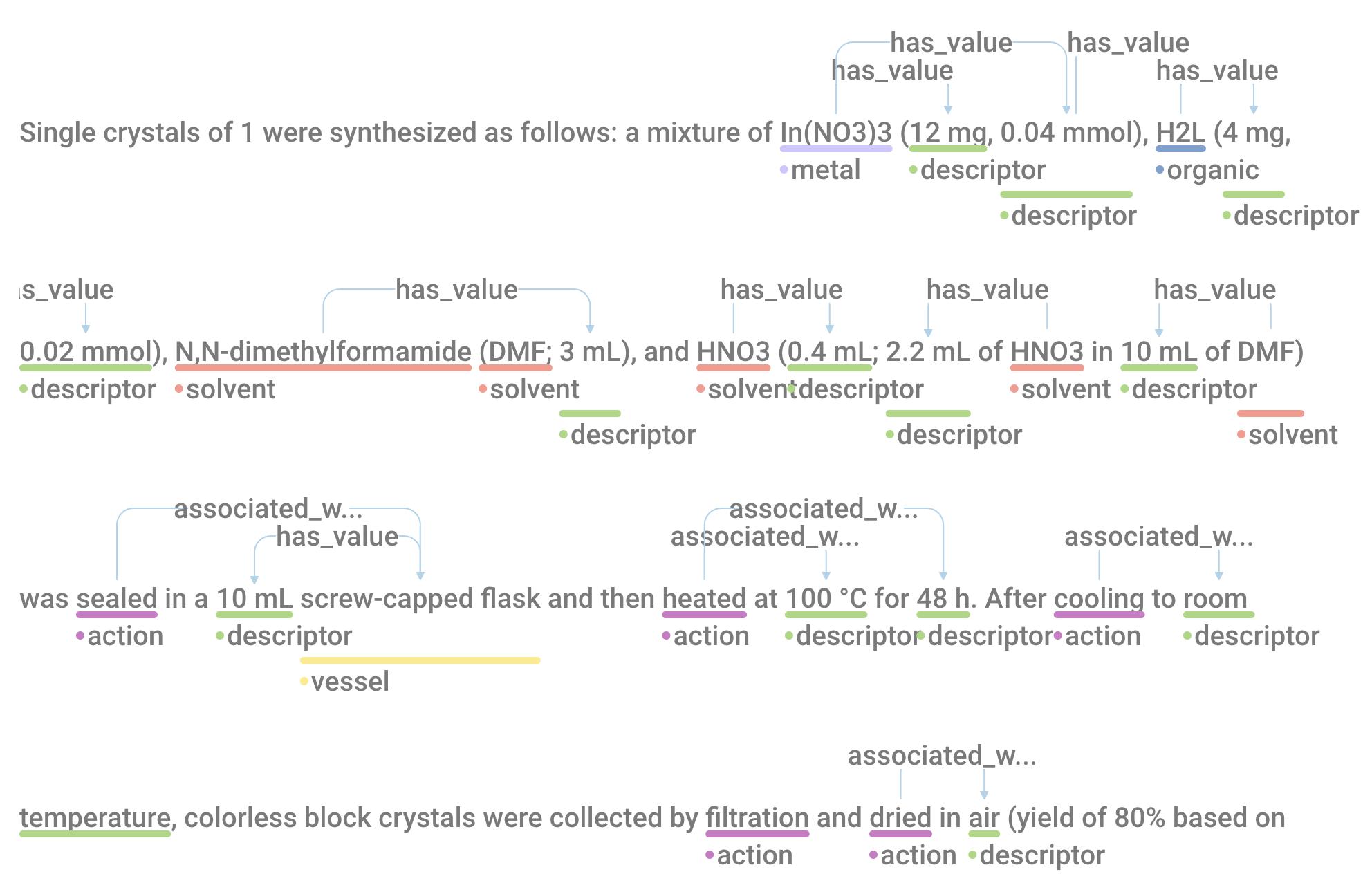}
\caption{An Example of Annotated Paragraph about MOF Synthesis based on our Metadata Schema} 
\label{fig2}
\end{figure}

During the annotation process, domain scientists recorded entity types that they believe worth incorporating into the schema. The annotation procedure was performed on a software tool called Doccano. Figure \ref{fig2} provides a visualized example of synthesis report annotation based on our metadata design.

\subsection{Further Modification based on Experts' Feedback}
We analyzed the 15 annotated paragraphs and arranged follow up discussion with domain scientists. Based on this first-round annotation, our domain scientists suggested that all components in the current metadata design are important and hence should be kept. However, the current metadata design is not sufficient to capture all critical information from a synthesis procedure. In this way, we incorporated new entity types suggested by scientists to our metadata schema. The following three entity types were added: 1) Vessel 2) MOF structure and 3) Acid.

\section{Results: Current Progress and Future Work}
Having observed these disparities in MOF experiment reporting above, we noticed an urge to standardize these unstructured yet very valuable synthesis reports. To this end, we address this need by providing the following three outputs: (1) a research framework for metadata design, (2) a metadata schema that includes nine entities and two relationships for reporting MOF synthesis experiments, and (3) a growing database of MOF synthesis reports structured by our metadata scheme.

\begin{figure}
\centering
\includegraphics[width=0.9\textwidth]{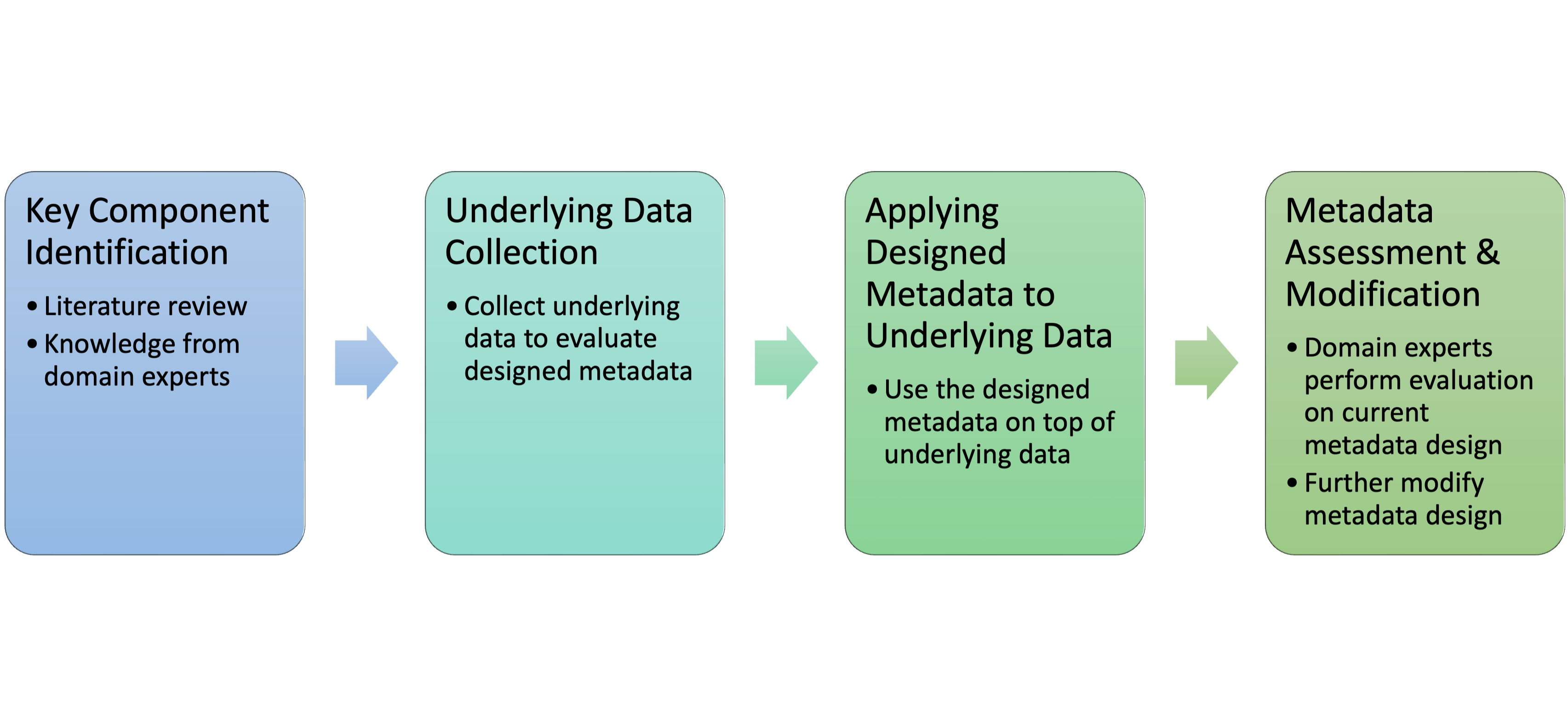}
\caption{A Generalized Framework for Designing Metadata for Scientific Experiments} 
\label{fig3}
\end{figure}

In order to better adapt our framework to broader scientific experiment areas, we generalized our proposed research framework in section 5 and proposed a general framework for experimental metadata design. As demonstrated in Figure \ref{fig3}, our research framework with human-in-the-loop can be generalized to four main steps: (1) key component identification, (2) underlying data collection, (3) applying designed metadata to the underlying data and (4) assessment on designed metadata and further modification.

While designing a metadata to describe experimental data, identifying key components in the target underlying data is the first step. The initial design of metadata for scientific experiments can be done by conducting literature review of previous established literature and obtaining knowledge from human experts. Then it is important to test our metadata design by mapping it to collected target experimental data. Finally, domain experts will jump in again at the final assessment of metadata design and make further modification on the design if needed.

\begin{figure}
\centering
\includegraphics[width=0.6\textwidth]{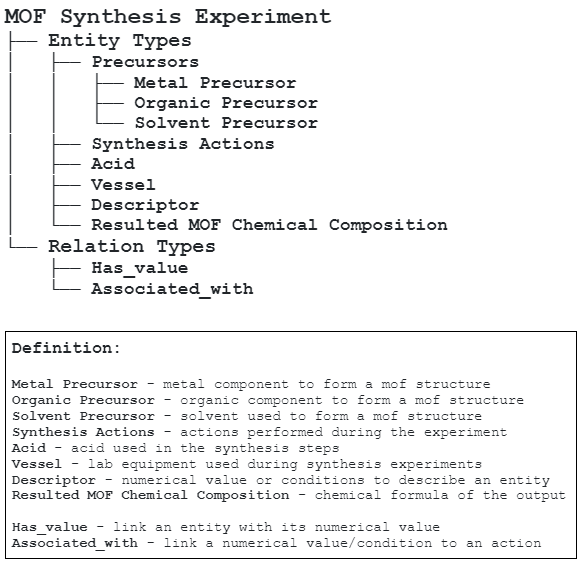}
\caption{Overall Metadata Design for MOF Synthesis Experiments} 
\label{tab1}
\end{figure}

As a result of the Metal-Organic Framework case study, our final design of metadata for MOF synthesis experiment is shown in figure \ref{tab1} as follows. Overall, we defined nine entities in hierarchical level: We have precursors, synthesis actions, acid, vessel, descriptor and resulted mof chemical composition as main entity types; the entity type “precursor” also has three following sub entity types: metal precursor, organic precursor and solvent precursor.

From May through August 2023, domain experts engaged in manually annotating 80 synthesis paragraphs based on the metadata schema reported in this paper. The annotated paragraphs are stored as a database of MOF materials synthesis experiments and each synthesized MOF structure is linked to its corresponding research article. In the near future, one work priority is to expand our synthesis database tagged by the above metadata standard. To apply our metadata design to a large amount of synthesis reports, we are conducting experiments on using generative AI models and 80 annotated reports as training data to perform automatic semantic labeling to the rest of 2000+ synthesis reports. 

\section{Discussion}
The need of structuring scientific experiment reports has gained increasing attention across various research communities \cite{electro,nandy,McLeroy}. There's even interest amongst publishers. This especially holds true for the metal-organic framework studies\cite{mofnlp,b25}: due to the complex design space of metal-organic framework materials, the way experimentalists describe their synthesis procedures significantly varies from each other. Other than the difference in writing style, they also tend to report different sets of aspects related to the synthesis experiment. As reported in the results, not all experimentalists report the type of vessels used in the synthesis procedure: some researchers believe vessels used during the experiment could make a difference in the final output, whereas some researchers do not think it would matter. Similarly, there is also a difference in the scope of reports, where some experiment reports are more detail-driven and some reports are more generalized.

Overall, this study on metadata design for MOF experimental data is beneficial to the broader research community – our framework for metadata design can be adapted to different scientific research areas.The growing MOF synthesis database structured by our metadata schema has high potential for multiple aspects, such as synthesis information retrieval, synthesis design and it ultimately helps the reproducibility of synthesis experiments.

This study is not without limitations. While initializing our metadata design for MOF experimental data, we interviewed four material scientists studying metal-organic framework structures. Three MOF experts were on the experimental side and one MOF expert was on the computational side.  However, there are many more scientists conducting research in the MOF area and we plan for further assessment of the initial entity set. In this way, this study can be limited in terms of the scope of human expert interviews.

\section{Conclusion}
In this paper, we identified the need to structure scientific experiment reports and conducted a case study on Metal-Organic Framework studies. We address this need by providing a research framework to design experimental metadata and a growing database for MOF synthesis reports structured by our designed metadata schema. The resulting metadata schema for MOF synthesis experiments can be beneficial to facilitate MOF material discovery. The database will be further expanded by leveraging large pre-trained language models for automatic semantic labeling in the short future.

%
%
%

\begin{thebibliography}{8}
\bibitem{b20}
Chao, T. C. (2014). Enhancing metadata for research methods in data curation. Proceedings of the American society for information science and technology, 51(1), 1-4.
\bibitem{b19}
Cowles, H. M. (2020). The scientific method: An evolution of thinking from Darwin to Dewey. Harvard University Press.
\bibitem{b8}
Darwin Core \url{https://www.tdwg.org/standards/dwc/}
\bibitem{b6}
DDI Alliance \url{https://ddialliance.org/Specification/}
\bibitem{b15}
Devaraju, A., Klump, J., Tey, V., Fraser, R., Cox, S., \& Wyborn, L. (2017). A digital repository for physical samples: Concepts, solutions and management. In Research and Advanced Technology for Digital Libraries: 21st International Conference on Theory and Practice of Digital Libraries, TPDL 2017, Thessaloniki, Greece, September 18-21, 2017, Proceedings 21 (pp. 74-85). Springer International Publishing.

\bibitem{dcc}
Digital Curation Centre \url{https://www.dcc.ac.uk/guidance/standards/metadata}

\bibitem{ecoinformatics}
Ecological Metadata Language \url{https://eml.ecoinformatics.org/}

\bibitem{Faundeen}
Faundeen, J. L., Burley, T. E., Carlino, J., Govoni, D. L., Henkel, H. S., Holl, S., ... \& Zolly, L. S. (2013). The United States geological survey science data lifecycle model. Reston, VA, USA: US Department of the Interior, US Geological Survey.

\bibitem{b9}
Fegraus, E. H., Andelman, S., Jones, M. B., \& Schildhauer, M. (2005). Maximizing the value of ecological data with structured metadata: an introduction to ecological metadata language (EML) and principles for metadata creation. Bulletin of the Ecological Society of America, 86(3), 158-168.


\bibitem{b13}
Food and Agricultural Organization of the United Nations \url{https://www.fao.org/faoterm/language-resources/en/}

\bibitem{b24}
Glasby, L. T., Gubsch, K., Bence, R., Oktavian, R., Isoko, K., Moosavi, S. M., ... \& Moghadam, P. Z. (2023). DigiMOF: A Database of Metal–Organic Framework Synthesis Information Generated via Text Mining. Chemistry of Materials.

\bibitem{b14}
Grabus, S., \& Greenberg, J. (2019). The landscape of rights and licensing initiatives for data sharing. Data Science Journal, 18, 29-29.

\bibitem{b22}
Huvila, I., Greenberg, J., Sköld, O., Thomer, A., Trace, C., \& Zhao, X. (2021). Documenting Information Processes and Practices: Paradata, Provenance Metadata, Life‐Cycles and Pipelines. Proceedings of the Association for Information Science and Technology, 58(1), 604-609.

\bibitem{b21}
Huvila, I., \& Sinnamon, L. (2022). Sharing Research Design, Methods and Process Information in and out of Academia. Proceedings of the Association for Information Science and Technology, 59(1), 132-144.

\bibitem{b18}
Katz, D. S., Gruenpeter, M., \& Honeyman, T. (2021). Taking a fresh look at FAIR for research software. Patterns, 2(3).

\bibitem{b23}
Leipzig, J., Nüst, D., Hoyt, C. T., Ram, K., \& Greenberg, J. (2021). The role of metadata in reproducible computational research. Patterns, 2(9).

\bibitem{b2}
McKiernan, E. C., Bourne, P. E., Brown, C. T., Buck, S., Kenall, A., Lin, J., ... \& Yarkoni, T. (2016). How open science helps researchers succeed. elife, 5, e16800.

\bibitem{McLeroy}
McLeroy, K. R., Garney, W., Mayo-Wilson, E.,\& Grant, S. (2016). Scientific reporting\: raising the standards. Health Education \& Behavior, 43(5), 501-508.

\bibitem{rdamsc}
Metadata Standards Catalog \url{https://rdamsc.bath.ac.uk/}

\bibitem{b16}
Michiorri, A., Sempreviva, A. M., Philipp, S., Perez-Lopez, P., Ferriere, A., \& Moser, D. (2022). Topic Taxonomy and Metadata to Support Renewable Energy Digitalisation. Energies, 15(24), 9531.

\bibitem{nandy}
Nandy, A., Duan, C., \& Kulik, H. J. (2021). Using machine learning and data mining to leverage community knowledge for the engineering of stable metal–organic frameworks. Journal of the American Chemical Society, 143(42), 17535-17547.

\bibitem{openai}
https://pypi.org/project/openai/

\bibitem{mofnlp}
Park, S., Kim, B., Choi, S., Boyd, P. G., Smit, B., \& Kim, J. (2018). Text mining metal–organic framework papers. Journal of chemical information and modeling, 58(2), 244-251.

\bibitem{b25}
Park, H., Kang, Y., Choe, W., \& Kim, J. (2022). Mining insights on metal–organic framework synthesis from scientific literature texts. Journal of Chemical Information and Modeling, 62(5), 1190-1198.

\bibitem{b5}
Rasmussen, K. B. (2014). Social Science Metadata and the Foundations of the DDI. IASSIST Quarterly, 37(1-4), 28-28.

\bibitem{b17}
Stocker, M., Darroch, L., Krahl, R., Habermann, T., Devaraju, A., Schwardmann, U., ... \& Häggström, I. (2020). Persistent identification of instruments. arXiv preprint arXiv:2003.12958.

\bibitem{b3}
Stodden, V. (2010). Open science: policy implications for the evolving phenomenon of user-led scientific innovation. Journal of Science Communication, 9(01), A05.


\bibitem{b12}
Subirats, I., Malapela, T., Dister, S., Zeng, M., Goovaerts, M., Pesce, V., ... \& Keizer, J. (2012). Reorienting open repositories to the challenges of the semantic web: experiences from FAO’s contribution to the resource processing and discovery cycle in repositories in the agricultural domain. In Metadata and Semantics Research: 6th Research Conference, MTSR 2012, Cádiz, Spain, November 28-30, 2012. Proceedings 6 (pp. 158-167). Springer Berlin Heidelberg.

\bibitem{electro}
Trench, B. (2009). Science reporting in the electronic embrace of the Internet. na.

\bibitem{b4}
Uhlir, P. F., \& Schröder, P. (2007). Open data for global science. Data Science Journal, 6, OD36-OD53.

\bibitem{b7}
Wieczorek, J., Bloom, D., Guralnick, R., Blum, S., Döring, M., Giovanni, R., ... \& Vieglais, D. (2012). Darwin Core: an evolving community-developed biodiversity data standard. PloS one, 7(1), e29715.

\bibitem{b1}
Wilkinson, M. D., Dumontier, M., Aalbersberg, I. J., Appleton, G., Axton, M., Baak, A., ... \& Mons, B. (2016). The FAIR Guiding Principles for scientific data management and stewardship. Scientific data, 3(1), 1-9.

\bibitem{usgs}
U.S. Geological Survey 
\newline \url{https://www.usgs.gov/data-management/metadata-creation}

\bibitem{yaghi2003}
Yaghi, O. M., O'Keeffe, M., Ockwig, N. W., Chae, H. K., Eddaoudi, M., \& Kim, J. (2003). Reticular synthesis and the design of new materials. Nature, 423(6941), 705-714.

\end{thebibliography}
%

\end{document}